\documentclass[preprint,showpacs,12pt]{revtex4-1} 
\usepackage{amsmath,amssymb,graphicx}
\begin{document}

\title{Intrinsic-Correlation Quantum Key Generation}

\author{Kim Fook Lee}

\affiliation{ Department of Electrical Engineering and Computer Science,\\ Northwestern University,\\ Evanston, Illinois 60208}

\author{Yong Meng Sua}

\affiliation{ Department of Physics,\\ Michigan Technological
University,\\ Houghton, Michigan 49931}

\begin{abstract}

A new conceptual key generation scheme is presented by using
intrinsic quantum correlations of single photons between Alice and
Bob. The intrinsic bi-partite correlation functions allow key bit to
be generated through high level communication language i.e. a key
bit is directly encoded to shared correlation functions not to the
state and detection of a photon at Bob does not mean key bit. These
make the scheme robust against intercept-resend attack because Alice
and Bob can always check the errors in their measurements and reveal
the presence of Eve in their channel without leaking any key bit
information. The scheme is strictly relied on the perfect
beam splitter and mean photon number less than 1, where more than
one photon in a coherent pulse will introduce more errors in Bob
even without the presence of Eve. From the percentage of errors in
Bob, we can estimate the amount of information will be leaked to Eve
in the photon-number splitting attack. This scheme can preserve the
randomness of phase-randomized light source for doubling the
communication distance as in original Ekert's protocol and providing
the raw key generation rate a factor of 2 higher than weak coherent
light protocols.

\end{abstract}

\maketitle 

\section{INTRODUCTION}

The quantum mechanics without probability amplitude was
proposed~\cite{wootters86}, leading to the possibility of quantum
information processing more directly in terms of probability i.e
discrete Wigner function~\cite{saraceno02}. In the same vein, it is
exciting to explore quantum key generation~\cite{Gisin02,Scarani09}
with polarization correlation function or expectation value of two
observers. In general, two observers $\mathcal{A}$ and $\mathcal{B}$
will have eigenvalues $\pm 1$ corresponding to photon detection at
the transmitted or reflected port of their analyzers. The
correlation function shared by them is $\mathcal{C}(\theta_a,
\theta_b)=\sum_{\pm 1}(\mathcal{A}\mathcal{B}) \mathcal{P}(\theta_a,
\theta_b)$, where $\mathcal{P}(\theta_a, \theta_b)$ is the joint
probability of detecting a photon in observer $\mathcal{A}$ and
$\mathcal{B}$ with their analyzers at $\theta_a$ and $\theta_b$,
respectively. The maximum/minimum values of $\mathcal{C}(\theta_a,
\theta_b)=\pm 1$ are corresponding to correlated/anti-correlated
outcomes of $\mathcal{A}$ and $\mathcal{B}$ such that the
$\mathcal{C}=+1 (\mathcal{A}=+1, \mathcal{B}=+1 ; \mathcal{A}=-1,
\mathcal{B}=-1)$ and the $\mathcal{C}=-1 (\mathcal{A}=+1,
\mathcal{B}=-1 ; \mathcal{A}=-1,\mathcal{B}=+1)$. There are totally
four correlation functions $\mathcal{C}_{n}=\pm Cos 2 (\theta_a \pm
\theta_b)$, where n = 1, 2, 3, 4, available from intrinsic
polarization correlation of weak coherent states~\cite{sua11},
coherent light mixed with random noise field~\cite{kflee09} and
coherent light fields~\cite{kflee02, kflee04}. These four correlation functions have been demonstrated and used to  perform bits correlations between two parties over a distance of 10 km, through a transmission fiber by using weak coherent states in Ref.5. One can predict the
outcome of the other by guessing the $\mathcal{C}_{n}$ through their
projection angles and photon detection. The secure key bits between
two observers $\mathcal{A}$ and $\mathcal{B}$ are then generated
through sharing four bi-partite correlation functions and photon
counting.

The essence of the paper is to propose a new scheme for key
generation based on single photons bi-partite correlation functions.
The proposed experiment setting is similar to simplified EPR
protocol (BBM92)~~\cite{BB92b}, where the EPR pairs are replaced
with two weak coherent states prepared from a light source in the
middle of Alice and Bob. The light source is two modes (x, y) weak
coherent states combined through a beam splitter producing two
spatially separated quantum channels. We have the product states at
the input of the beam splitter such that $|\alpha\rangle_{x}(\pm
i|\beta\rangle_{y})=e^{-\frac{n_{x}}{2}}e^{-\frac{n_{y}}{2}}\sum^{\infty}_{n=0}
\sum^{\infty}_{m=0}
\frac{n_{x}^{n}}{\sqrt{n!}}\frac{n_{y}^{m}}{\sqrt{m!}}|n, m\rangle
\propto |0\rangle_{x}|0\rangle_{y}\pm i
\sqrt{n_{x}}|1\rangle_{x}|0\rangle_{y}\pm i \sqrt{n_{y}}
|0\rangle_{x}|1\rangle_{y}+..$, where $\pm i$ is provided by random
phase modulator $\pm \frac{\pi}{2}$ along the y-direction and
{$n_{x}, n_{y}\leq 1$}. The density matrix of the output state from the beam splitter is dependant on the integration over the phase space of P representation for the input product state. Since the $|\alpha\rangle_{x}$ and $|\beta\rangle_{y}$ are intrinsically correlated from the same laser, the integration over the phase spaces of $|\alpha\rangle_{x}$ and $|\beta\rangle_{y}$ can produce the output state that is intrinsically entangled~\cite{Schleich01}. One can see that, the beam splitter transforms the input state $|\alpha\rangle_{x}(\pm
i|\beta\rangle_{y})$ into inseparable output state $(\frac{|\alpha\rangle_{x}+
ie^{\pm
i\phi_{m}}|\beta\rangle_{y}}{\sqrt{2}})_{2}(\frac{i|\alpha\rangle_{x}+e^{\pm
i\phi_{m}}|\beta\rangle_{y}}{\sqrt{2}})_{1}$, where $\phi_{m}=\pm 90^{\circ}$ is random modulated phase. The  inseparable output state can provide four types of bi-partite correlation functions $(\mathcal{C}_{1,2,3,4})$ by manipulating the linear phase shifters in Box 1 of Alice. The security of this scheme is guaranteed by the fact that the nonorthogonality of the product coherent states,
which cannot be identified by a single measurement. The protocol is
strictly depended on equal mean photon number in the mode (x, y)
i.e. $n_{x}=n_{y}=n_{c}< 1$. We will discuss how this condition can
restrict more than one photon in a coherent pulse in this protocol
and hence to prevent the photon-number splitting attack. The single
photons sent to Alice and Bob are called ancilla photon and signal
photon, respectively. The scheme requires coincidence detection of
ancilla photon at Alice and signal photon at Bob. The ancilla photon
passes through/reflects from an analyzer at Alice is registered as
bit '1'/'0'. The signal photon passes through/reflects from an
analyzer at Bob is denoted as 'yes'/'no' of his guess on Alice's key
bit. 'Yes' did not mean that Bob will have bit '1'. It can be bit
'0'. Alice and Bob generate key bits through secure communication
 in the sense that the bit information is encoded on the
$\{\mathcal{C}_{n}\}$ not the state. We will discuss how the
protocol is robust against intercept-resend attack. The security
analysis of this scheme may well be different from current weak
light
protocols~\cite{Brassard00,Lutkenhaus00,Hwang03,Lo05,Wang05,Scarani04}.

In this paper, we first outline the sequence of steps for
implementing key generation between Alice and Bob. Then, we discuss
the physics and experimental detail of each step. Finally, we
briefly discuss how the protocol can prevent the photon-number
splitting and intercept-resend attacks.

\section{THE IC-QKD PROTOCOL}
The proposed experiment setup is shown in Fig.1. The setting is
similar to the BBM92 protocol, where the EPR pairs are replaced with
two mode and phase-modulated weak coherent states, which are
intrinsic-correlated from a laser source in the middle of Alice and
Bob. The two weak coherent states are distributed through two pulses
within a bit period, which is phase modulated in an asymmetric
Mach-Zehnder interferometer (MZI).

\begin{figure}[htbp]
\centerline{\includegraphics[width=.8\columnwidth]{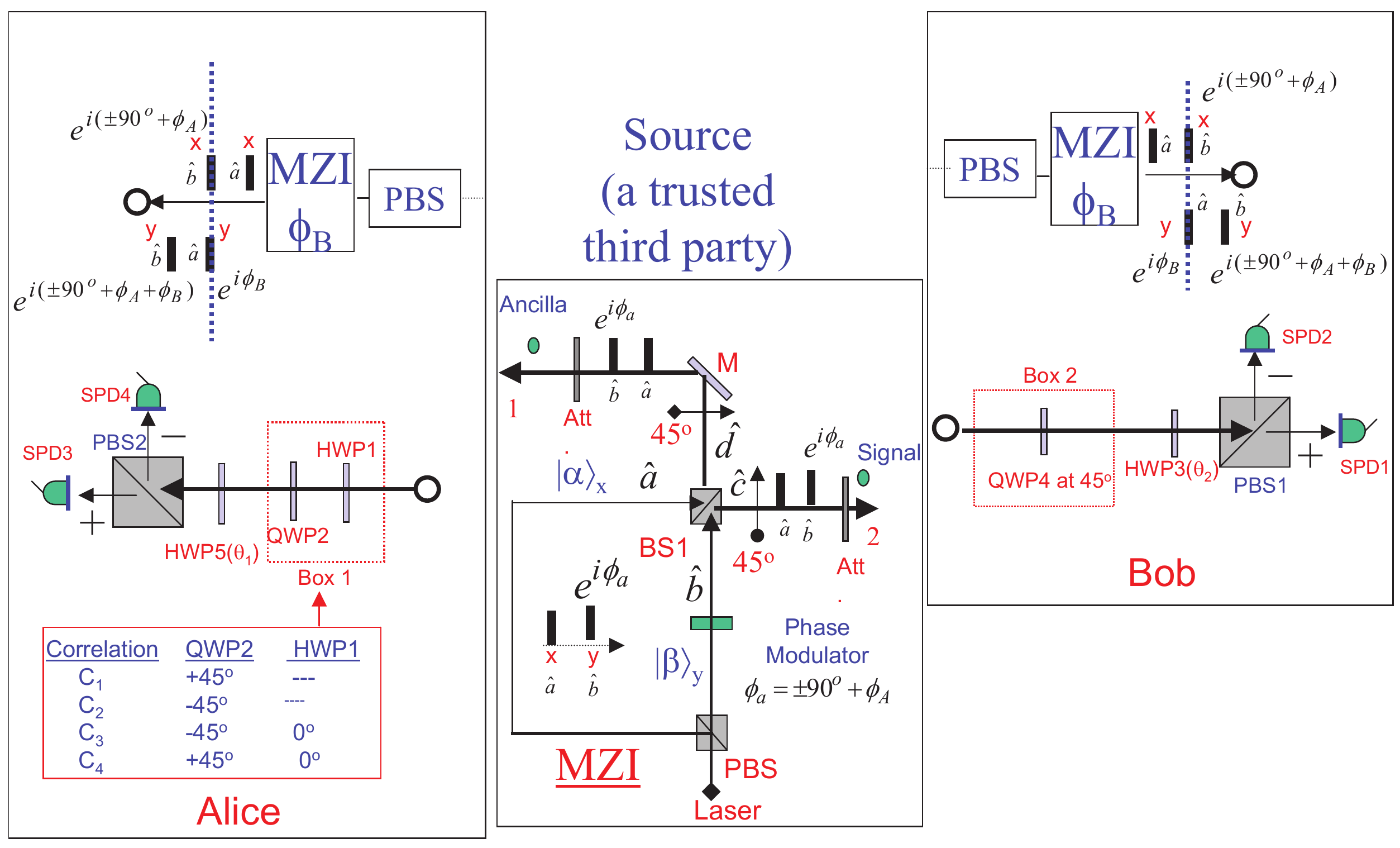}}
\caption{The proposed experiment setup for implementing key
generation using a signal photon at Bob and an ancilla photon at
Alice. The laser source is located in the middle of Alice and Bob,
and secured by a trusted third party. The two weak coherent states
are distributed through two pulses within a bit period. The dotted
boxes (Box 1 and Box 2) are the wave plates used for preparing
bi-partite correlation between Alice and Bob. HWP: Half-Wave Plate,
QWP: Quarter-Wave Plate, SPD: Single Photon Detector, PBS:
Polarization Beam Splitter, BS: Beam Splitter.}
\end{figure}

Four bi-partite correlation functions $(\mathcal{C}_{n}, n=1,2,3,4)$
are assigned into two groups $(\Psi, \Phi)$, where
$\mathcal{C}_{1,2} \rightarrow \Psi$ and $\mathcal{C}_{3,4}
\rightarrow \Phi$. Alice will have the key bit '0' and '1' for the
correlation functions $\mathcal{C}_{1,3}$ and $\mathcal{C}_{2,4}$,
respectively. Alice can randomly choose the correlation function
$\{\mathcal{C}_{n}\}$ and then her key bits. Bob guesses the
correlation function based on his polarization angle of his analyzer
and records the outcome of his measurement i.e. 'yes' or 'no' of his
guess. After the signal transmission, Alice then tells Bob through
classical channel about the sequence of groups $\{\Psi, \Phi\}$ that
she has randomly chosen. Bob can then know Alice's key bit by the
outcome of his measurement regardless he made a right or wrong
guess, so doubling the raw key generation rate.

\begin{figure}[htbp]
\centerline{\includegraphics[width=.8\columnwidth]{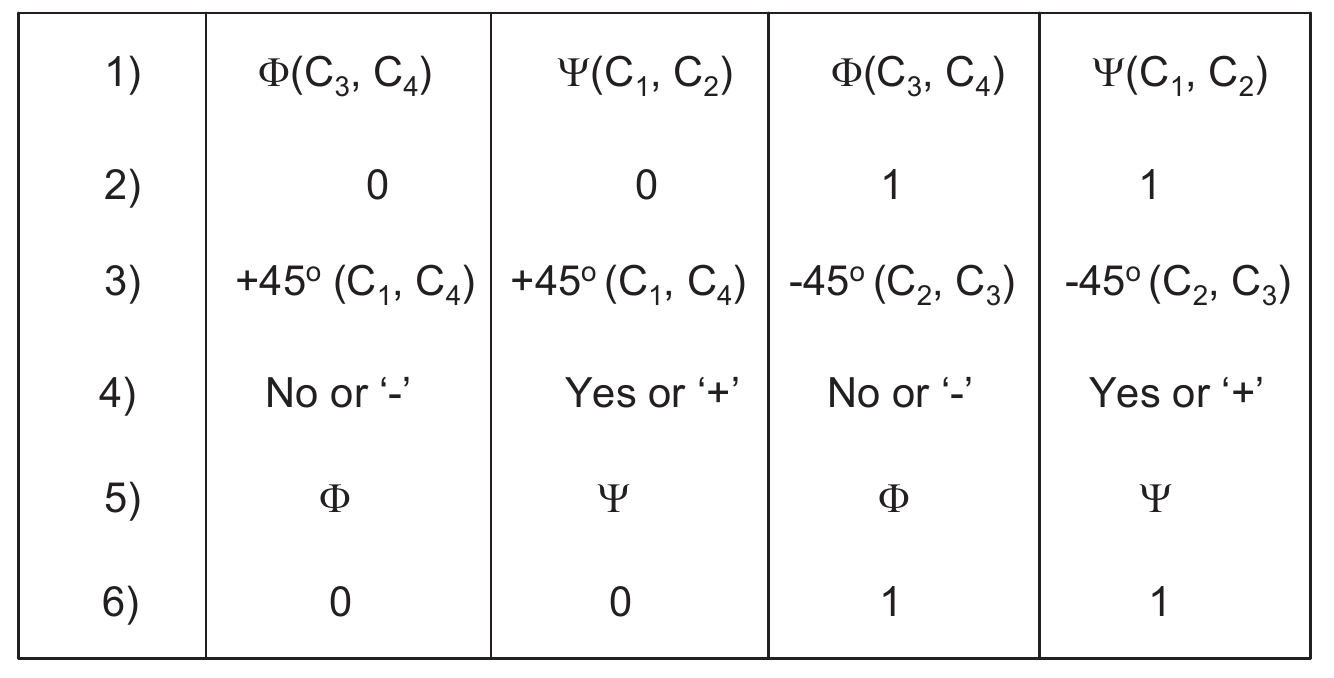}}
\caption{The scheme for key generation between Alice and Bob.}
\end{figure}

To illustrate the scheme more systematically, we will discuss an
example of the key generation as shown in Fig.2.

$\underline{Step-1}$: Alice randomly chooses the
$\Psi\{\mathcal{C}_{1}, \mathcal{C}_{2}\}$ or
$\Phi\{\mathcal{C}_{3}, \mathcal{C}_{4}\}$ by choosing the
waveplates combination in Box 1 as shown in Fig.1. (Alice randomly
chooses the $\mathcal{C}_{n}$ is analog to choosing a random basis
in BB84.)

$\underline{Step-2}$: Alice fixes her polarizer at angle
$\theta_{1}=+45^{\circ}$ and records the valid detections of the
ancilla photon in '+' SPD3 and '-' SPD4 as bit '1' or '0'.

$\underline{Step-3}$: Bob guesses on the {$\mathcal{C}_{n}$} by
projecting his polarizer at angles $\theta_{2}=\pm 45^{\circ}$. The
$\theta_{2}=+45^{\circ}$ means Bob's guess on $\{\mathcal{C}_{1},
\mathcal{C}_{4}\}$. The $\theta_{2}=-45^{\circ}$ means Bob's guess
on $\{\mathcal{C}_{2}, \mathcal{C}_{3}\}$.

(Alice only chooses one polarization angle $\theta_{1}=+45^{\circ}$
and Bob chooses two polarization angles $\theta_{2}=\pm 45^{\circ}$
because of the shared $\mathcal{C}_{n}$. We do not need to perform
the violation of Bell inequality in Ekert's protocol. And also, not
like BBM92, Alice and Bob do not need to measure the basis
$(0^{\circ}$, $90^{\circ})$ of the incoming photon.)

$\underline{Step-4}$: The outcomes of Bob's guess through the valid
detections of signal photon. 'Yes' means that the'+' SPD1 'click' or
his guess is right. 'No' means that the '-' SPD2 'click' or his
guess is wrong. However, Bob did not know the Alice's key bit yet.

(Alternative: Alice and Bob can detect the presence of Eve by
broadcasting part of the results obtained by Bob in step 3 and 4.
They did not leak any bit information to Eve because Alice will not
announces the group information in step 5 for the exposed data. Note
that in BB84, Alice and Bob announce their choice of bases, and Bob
never broadcast his result which is the key bit.)

$\underline{Step-5}$: After the signal transmission, Alice announces
to Bob through classical channel about the group of her choice
$(\Phi, \Psi)$, not revealing her choice of correlation function.

$\underline{Step-6}$: Bob knows Alice's key bit after he received
the group information and verified with the record of his valid
detection. To illustrate this, let's say for the first column, Alice
announces to Bob that the group is $\Phi$. Bob's guess on the group
$\Phi$ is $\mathcal{C}_{4}$, which is corresponding to bit '1'.
However, he knows that the guess is wrong through his valid
detection in step 4. From here, he knows Alice's key bit is bit '0'.

(The random choices of $\mathcal{C}_{n}$ in Alice and projection
angle $\theta_{2}=\pm 45^{\circ}$ in Bob did not need to discard
half of the raw keys compared to BB84 where Alice and Bob have the
wrong bases.)

We believe that this scheme can provide the raw key generation rate
a factor of 2 higher than the BB84~\cite{BB84} and B92~\cite{BB92a},
and double the communication distance as in the simplified EPR
(BBM92)~\cite{BB92b} and the original Ekert's
protocol~\cite{Ekert91}. We would outline the detail explanation of
each step.

\subsection{Step 1, 2: Prepare and Measure}

The light source is located in the middle of Alice and Bob. A
pulsed, $45^{\circ}$-polarized laser light with a polarizing beam
splitter (PBS) is used to provide a weak coherent state $|\alpha
\rangle_{x}$ with horizontal polarization and another weak coherent
state $|\beta\rangle_{y}$ with vertical polarization in an
asymmetric Mach-Zehnder interferometer (MZI). These coherent states
are located at two time slots and combined through a perfect 50/50
beam splitter (BS1), producing two spatially separated beams, i.e.,
beam 1 and beam 2. The input state of the beam splitter is a product
state of two mode weak coherent states $|\alpha\rangle_{x}|\beta
\rangle_{y}$, which is the prepared state for this protocol. The
annihilation operators $\hat{a}$ and $\hat{b}$ are the input field
operators for the beam splitter (BS1) and for the coherent states
$|\alpha\rangle_{x}$ and $|\beta\rangle_{y}$. The operator $\hat{b}$
is phase-modulated with the phase $\pm \phi_{a}$ as $\hat{b}e^{\pm i
\phi_{a}}$, where $\phi_{a}= \phi_{m}+\phi_{A}$, and $\phi_{m}=
\pm 90^{\circ}$ is randomly chosen. The $\phi_{a}$ is produced by a
phase modulator at the short arm of the MZI. The annihilation
operators
$\hat{c}=\frac{1}{\sqrt{2}}(\hat{a}\hat{x}+i\hat{b}\hat{y})$ and
$\hat{d}=\frac{1}{\sqrt{2}}(i\hat{a}\hat{x}+\hat{b}\hat{y})$ are the
output field operators for the BS1, which are sent through beam 2 to
Bob and beam 1 to Alice, respectively. To distribute the operators
$\hat{d}$ and $\hat{c}$ to Alice and Bob, we first place a polarizer
at $45^{\circ}$ at each output of the BS1 so that two pulses are
co-polarized within a bit period. The reason is two orthogonal
polarized pulses may be vulnerable to Eve attacks. To create single
photon quantum channel between the source and Bob, the beam 2 is
attenuated to single photon level with the mean photon number per
pulse $(n_{x}, n_{y}) \leq 1$. Similarly, the beam 1 is further
attenuated to single photon level with mean photon number per pulse
$(n_{x}, n_{y}) \leq 1$. The co-polarized two pulses are sent
through transmission fibers to Alice's and Bob's MZI, where the
phase demodulators at the short arms induced a phase of
$\theta_{B}$. Four pulses ( 2 in the x-mode and 2 in the y-mode) are
created within one bit period. The x-y mode ($e^{i(\pm
90^{\circ}+\phi_{A})}, e^{i\phi_{B}}$) in the middle time slot are
used for key generation. The phase ($\phi_{A}-\phi_{B}=0$) can be
accomplished by making use of the phase coding $\alpha\eta$
system\cite{Barbosa03,Corndorf03,Corndorf05,Liang05}, which is
proven to be more efficient secure encryption system over long
distances. Note that we apply phase coding $\alpha\eta$ system on a
single photon level instead of mesoscopic coherent state.

The random phase $\phi_{m}=\pm 90^{\circ}$ is unknown to Bob and
Alice. With the assumption of low mean photon numbers for $n_{x}$
and $n_{y}$, the two weak coherent states can be treated as
$|\tilde{\alpha}\rangle_{x}|\tilde{\beta}\rangle_{y} =
e^{\frac{1}{2}(n_{x}+n_{y}}[|0\rangle_{x}|0\rangle_{y}
+\sqrt{n_{x}}|1\rangle_{x}|0\rangle_{y}+\sqrt{n_{y}}|0\rangle_{x}|1\rangle_{y}....]$,
where we have considered the vacuum and one photon number state in
mode x or y. The phase $\phi_{m}=\pm 90^{\circ}$ is not included
here because we have treated the $\phi_{m}$ in the annihilation
operator $\hat{b}e^{i (\pm\phi_{m}+\phi_{A})}$. Four types of
intrinsic bi-partite correlation functions $\mathcal{C}_{1,2}=\{
-cos 2(\theta_{1}- \theta_{2}), +cos 2(\theta_{1}+\theta_{2})\}$ and
$\mathcal{C}_{3,4} = \{ -cos 2(\theta_{1}+ \theta_{2}), +cos
2(\theta_{1}-\theta_{2})\}$ can be established through the
combination of wave plates as shown in the dotted boxes (Box 1 and
Box 2) in Fig.1. The half-wave plates (HWP3 and HWP5) before the
polarizing beam splitters (PBS1 and PBS2) at Bob and Alice are used
for projecting polarization angles $\theta_{1}$ and $\theta_{2}$,
respectively, so that the maximum correlation
$\mathcal{C}_{1,2,3,4}= \pm 1$ is obtained.

The annihilation operator $\hat{d}$ at beam 1 is transformed through
the combination of waveplates in Box 1 and a polarization analyzer
consists of a half waveplate (HWP5) and a polarizing beam splitter
(PBS2). Alice can randomly choose the $\mathcal{C}_{n}$, n=1,2,3,4
through the settings of QWP2 and HWP1 as shown in the Box 1. For the
correlation functions $\mathcal{C}_{1,2,3,4}$, the photon number
operator $\hat{d'}^{\dagger}\hat{d'}$ at Alice is given by,
\begin{eqnarray}
\hat{d'}^{\dagger}\hat{d'}=\frac{1}{2}[\hat{a}^{\dagger}\hat{a}+\hat{b}^{\dagger}\hat{b}
\pm \hat{a}^{\dagger}\hat{b}e^{\pm i 2 \theta_{1}+ i \phi_{m}}+
c.c]. \label{eq:01}
\end{eqnarray}
The expectation values for the correlation $\{\mathcal{C}_{1,2}\}$
are given by,
\begin{eqnarray}
 \mathcal{C}_{1}\rightarrow \langle \tilde{\alpha \beta}|
\hat{d'}^{\dagger}\hat{d'}|\tilde{\alpha \beta}\rangle \propto
[n_{x}+n_{y}- 2\sqrt{n_{x}n_{y}}cos (2\theta_{1}+
\phi_{m})]\nonumber\\
\mathcal{C}_{2}\rightarrow \langle \tilde{\alpha \beta}|
\hat{d'}^{\dagger}\hat{d'}|\tilde{\alpha \beta}\rangle \propto
[n_{x}+n_{y}+ 2\sqrt{n_{x}n_{y}}cos (2\theta_{1}- \phi_{m})],
\label{eq:02}
\end{eqnarray}
respectively. While the expectation values for the correlation
$\{\mathcal{C}_{3,4}\}$ are given by,
\begin{eqnarray}
\mathcal{C}_{3}\rightarrow \langle \tilde{\alpha \beta}|
\hat{d'}^{\dagger}\hat{d'}|\tilde{\alpha \beta}\rangle \propto
[n_{x}+n_{y}- 2\sqrt{n_{x}n_{y}}cos (2\theta_{1}-
\phi_{m})]\nonumber\\
\mathcal{C}_{4}\rightarrow \langle \tilde{\alpha \beta}|
\hat{d'}^{\dagger}\hat{d'}|\tilde{\alpha \beta}\rangle
\propto[n_{x}+n_{y}+ 2\sqrt{n_{x}n_{y}}cos (2\theta_{1}+ \phi_{m})],
\label{eq:03}
\end{eqnarray}
respectively. We have neglected the $\frac{1}{2}e^{-(n_{x}+n_{y})}$
and higher photon number states for the simplicity of the following
discussions. Note that the condition $n_{x}=n_{y}$ (perfect beam
splitter) has made the higher photon number states having the same
form of interference terms. Alice will observe the interference term
of $- Cos (2 \theta_{1} + \phi_{m})$ for the $\mathcal{C}_{1}$, $+
Cos (2 \theta_{1} - \phi_{m})$ for the $\mathcal{C}_{2}$, $- Cos (2
\theta_{1} - \phi_{m})$ for the $\mathcal{C}_{3}$, and $+ Cos (2
\theta_{1} + \phi_{m})$ for the $\mathcal{C}_{4}$. For example, if
Alice chooses $\mathcal{C}_{1}$, she has the interference term of $-
Cos (2 \theta_{1} + \phi_{m})$. For the mean photon number per pulse
$n_{x}=n_{y}=n_{c} \leq 1.0$. The expectation value $\langle
\hat{d'}^{\dagger}\hat{d'}\rangle$ will have the maximum value of
$2n_{c}e^{-2n_{c}}$ and minimum value of 0. The maximum value of
$\langle \hat{d'}^{\dagger}\hat{d'}\rangle$ is corresponding to the
interference term with value +1 and the minimum value (0) of
$\langle \hat{d'}^{\dagger}\hat{d'}\rangle$ is corresponding to the
interference term with value -1. In other words, the interference
term with value +1 indicates that the ancilla photon passed through
a PBS2. While the interference term with value -1 indicates that the
ancilla photon reflected from a PBS2. The interference term with the
value +1 (-1) is corresponding to the detection of the ancilla
photon at Alice's detector '+'SPD3 ('-'SPD4) or bit '1' (bit '0').
Note that we are not going to multiply Alice's and Bob's
interference term and measure the mean value of the multiplied
interference intensity as previously demonstrated in
ref~\cite{sua11}, i.e., $\mathcal{C}_{1,2,3,4}=\pm Cos 2
(\theta_{1}\pm \theta_{2})$ , where the phase term $\phi_{m}$ is
varied from $0^{\circ}\rightarrow 2\pi$.

\subsection{Step 3, 4: Guess and Verify}

The annihilation operator $\hat{c}$ at beam 2 is transformed through
a quarter waveplate (QWP4) at $45^{\circ}$ and a polarization
analyzer (HWP3 and PBS1). The photon number operator at Bob is given
by,
\begin{eqnarray}
\hat{c'}^{\dagger}\hat{c'}=\frac{1}{2}[\hat{a}^{\dagger}\hat{a}+\hat{b}^{\dagger}\hat{b}+\hat{a}^{\dagger}\hat{b}e^{i
2 \theta_{2}+ i \phi_{m}}+ \hat{b}^{\dagger}\hat{a}e^{- i 2
\theta_{2} - i \phi_{m}}],\label{eq:04}
\end{eqnarray}
and then the expectation value is given by
\begin{eqnarray}
\langle \tilde{\alpha}\tilde{\beta}|
\hat{c'}^{\dagger}\hat{c'}|\tilde{\alpha}\tilde{\beta}\rangle
\propto [n_{x}+n_{y}+ 2\sqrt{n_{x}n_{y}}cos
(2\theta_{2}+\phi_{m})].\label{eq:05}
\end{eqnarray}
Bob has the same interference term $cos (2 \theta_{2}+\phi_{m})$ for
all the {$\mathcal{C}_{n}$} because the QWP4 is projected at the
fixed angle $+45^{\circ}$ in Box 2. At Bob, if the single photon
detector (SPD) at '+' or '-' port detects the signal photon, he then
assigned the valid detection as 'Yes' or 'No' for his guesses on
$\mathcal{C}_{n}, n=1,2,3,4$.

In this scheme, the $\theta_{1}=+45^{\circ}$ at Alice is fixed.
Let's discuss the ideal situation where Alice chooses the
$\{\mathcal{C}_{n}\}$ with the right $\phi_{m}=\pm 90^{\circ}$ and
Bob guesses the right $\{\mathcal{C}_{n}\}$ with his projection
angle $\theta_{2}$. For example, if Alice chooses
$\{\mathcal{C}_{1}; \theta_{1}= +45^{\circ}\}$ with the $\phi_{m}=
-90^{\circ}$, the interference term in Alice has the value -1
indicating her '-' SPD4 will �click�. She records the detection as
bit '0'. While interference term in Bob with his projection angle
$\theta_{2}=+45^{\circ}$ has the value +1 indicating his '+' SPD1
will 'click'('yes'). This means that his guess on $C_{1}$ is right.
If Alice chooses $\{\mathcal{C}_{2}, \theta_{1}= +45^{\circ}\}$ with
the $\phi_{m}= +90^{\circ}$, her interference term indicates that
her '+' SPD3 will 'click' or bit '1' is registered. The interference
term in Bob with the $\theta_{2}=-45^{\circ}$ has the value +1 or
his '+' SPD1 will 'click'('yes'). This means that his guess on
$C_{2}$ is right. Similarly, if Alice chooses $\{\mathcal{C}_{3},
\theta_{1}= +45^{\circ}\}$ with the $\phi_{m}= +90^{\circ}$ and
$\{\mathcal{C}_{4}, \theta_{1}= +45^{\circ}\}$ with the $\phi_{m}=
-90^{\circ}$, she will register bit '0' and '1', respectively. Bob
will have the right guess on $\mathcal{C}_{3,4}$ with his projection
angles $\theta_{2}=+45^{\circ}, -45^{\circ}$, respectively.

We formulated the above discussion by assigning the
${\mathcal{C}_{n}}$ into two groups i.e.
$\mathcal{C}_{1,2}\rightarrow \Psi$ and
$\mathcal{C}_{3,4}\rightarrow \Phi$. Then, we can denote the groups
$\{\Psi, \Phi\}$ as a function of the $\mathcal{C}_{n}$, projection
angle $\theta$, and random phase $\phi_{m}=\pm 90^{\circ}$ for Alice
and Bob to generate the shared key bits. If Alice and Bob share the
group $\Psi\{\mathcal{C}_{1},\theta_{1}=+45^{\circ},
\theta_{2}=+45^{\circ}, \phi_{m}=-90^{\circ}\}$ which is
corresponding to bit '0', then the '-' SPD4 in Alice will detect the
ancilla photon and Bob's '+' SPD1 will detect the signal photon
meaning 'yes'. Similarly, for
$\Psi\{\mathcal{C}_{2},\theta_{1}=+45^{\circ},
\theta_{2}=-45^{\circ}, \phi_{m}=+90^{\circ}\} \rightarrow
\rm{bit}'1'$,  $\Phi\{\mathcal{C}_{3},\theta_{1}=+45^{\circ},
\theta_{2}=+45^{\circ}, \phi_{m}=-90^{\circ}\} \rightarrow
\rm{bit}'0'$,  $\Phi\{\mathcal{C}_{4},\theta_{1}=+45^{\circ},
\theta_{2}=-45^{\circ}, \phi_{m}=-90^{\circ}\} \rightarrow
\rm{bit}'1'$.

Since each group $\{\Psi, \Phi\}$ has bit '0' and '1' associated
with its correlation function $\mathcal{C}_{n}$ and depended on
random phase $\phi_{m}=\pm 90^{\circ}$, then how Alice knows her
choice of {$\mathcal{C}_{n}$} and valid detection in her detectors
can generate key bit in agreement with Bob's guess on
{$\mathcal{C}_{n}$} through his projection angle and valid detection
in his detectors. The beauty of this scheme is Alice and Bob can
generate the shared key bits through the shared correlation
functions without requiring any information about the random phase
$\phi_{m}=\pm 90^{\circ}$ in the source located in the middle
between them. To understand this more clearly, let's consider the
case where the phase $\phi_{m}=-90^{\circ}$ in the source. Alice
chooses the $\Psi\{\mathcal{C}_{1}, \theta_{1}=+45^{\circ},
\phi_{m}=-90^{\circ}\}\rightarrow \rm{bit}'0'$, Alice has the
interference term $- cos (2 \theta_{1}+\phi_{m}) \rightarrow - cos
(2 (45^{\circ})-90^{\circ})\rightarrow -1$ or '-' SPD4 will click
registering bit '0'. Bob has no idea about what Alice's choice on
the $\{\mathcal{C}_{n}\}$ and the phase $\phi_{m}=-90^{\circ}$. Bob
has the interference term $cos (2 \theta_{2}+\phi_{m})$. Bob can
only project the HWP3 to $\theta_{2}=\pm 45^{\circ}$. Bob chooses
the $\theta_{2}=+45^{\circ}$($\theta_{2}=-45^{\circ}$) for his guess
on $\Psi\{\mathcal{C}_{1}\}\rightarrow \rm{bit}'0'$ or
$\Phi\{\mathcal{C}_{4}\}\rightarrow \rm{bit}'1'$
($\Psi\{\mathcal{C}_{2}\}\rightarrow \rm{bit}'1'$ or
$\Phi\{\mathcal{C}_{3}\}\rightarrow \rm{bit}'0'$). Bob knows his
guess is right or wrong through the valid detection in his detector
'+' SPD1 or '-' SPD2, respectively. Bob did not know the key bit
until Alice announces which group $\{\Psi, \Phi\}$ in Step 5. Let's
say, Bob chooses $\theta_{2}=+45^{\circ}$, the interference term
$cos (2 (+45^{\circ}) -90^{\circ}) \rightarrow +1$ meaning his '+'
SPD1 will 'click'('yes') or his guess on $\Psi\{\mathcal{C}_{1}\}
\rightarrow '0'$ or $\Phi\{\mathcal{C}_{4}\}\rightarrow '1'$ is
right. After the signal transmission, Alice announces the group
$\Psi$ of her choice (Step 5). Bob knows the $\mathcal{C}_{1}$ is
right. Then, Alice and Bob  share the key bit '0' (Step 6). If Bob
chooses $\theta_{2}=-45^{\circ}$, the interference term $cos (2
(-45^{\circ}) -90^{\circ}) \rightarrow -1$ meaning his '-' SPD2 will
'click' or his guess on $\Psi\{\mathcal{C}_{2}\} \rightarrow '1'$ or
$\Phi\{\mathcal{C}_{3}\}\rightarrow '0'$ is wrong. After the signal
transmission, Alice announces the group $\Psi$ of her choice (Step
5). Bob knew that his guess on $\Psi\{\mathcal{C}_{2}\}$ or bit '1'
is wrong. Bob knows the bit is '0', then Alice and Bob will share
the key bit '0' (Step 6). Bob knows the key bits of Alice regardless
his guess is right or wrong. The above scenario is Alice's choice of
{$\mathcal{C}_{1}$} with right phase $\phi_{m}=-90^{\circ}$. The
most intriguing part of this scheme is Alice knows her key bit
regardless her choice of $\mathcal{C}_{n}$ in agreement with the
phase $\phi_{m}=\pm 90^{\circ}$. To discuss this, let's consider the
case where the phase $\phi_{m}=+90^{\circ}$, Alice still chooses the
$\Psi\{\mathcal{C}_{1}, \theta_{1}=+45^{\circ},
\phi_{m}=-90^{\circ}\}$ or bit '0', the interference term in Alice
$-cos (2 (45^{\circ})+90^{\circ})\rightarrow +1$ meaning that the
'+' SPD3 will 'click' or bit '1'. Alice knew then her choice of
$\mathcal{C}_{1}$ is wrong. She also knew that the phase
$\phi_{m}=+90^{\circ}$ was meant for the $\Psi\{\mathcal{C}_{2},
\theta_{1}=+45^{\circ}, \phi_{m}=+90^{\circ}\}\rightarrow
\rm{bit}'1'$, so she records the bit '1' as registered by her valid
detection. Amazingly, Bob will have the bit '1'. As discussed above,
if Bob chooses $\theta_{2}=+45^{\circ}$, the interference term $+
cos (2 (+45^{\circ}) +90^{\circ})\rightarrow -1$ meaning that his
'-' SPD2 will 'click' or 'No' on his guess. After Alice announces
the group $\Psi$, Bob knew that his guess on $\mathcal{C}_{1}$ or
bit '0' is wrong, so he knows that Alice has bit '1'. The similar
description is applied for the other $\Psi\{\mathcal{C}_{2}\}$,
$\Phi\{\mathcal{C}_{3}\}$, and $\Phi\{\mathcal{C}_{4}\}$.

\section{DISCUSSION}

Weak coherent state is usually vulnerable to photon-number splitting
attack i.e Eve can tap one signal photon from beam 2 and perform the
same measurement as Bob did. In this protocol, we make use of the
phase coding $\alpha\eta$
system~\cite{Barbosa03,Corndorf03,Corndorf05,Liang05} to secure beam
1 and beam 2 from Eve attacks. One can also use differential phase
shifted method to distribute operators $\hat{c}$ and $\hat{d}$ at
the output of the BS1. The protocol can prevent more than one photon
in a coherent pulse by applying the condition $n_{x}=n_{y}=n_{c}<
1$. For example, 2 photons in beam 2 can be attributed to higher
photon number in (x, y) mode such that $|1\rangle_{x}|1\rangle_{y},
|0\rangle_{x}|2\rangle_{y}, |2\rangle_{x}|0\rangle_{y}$, which will
have the expectation value of
$\langle\hat{c'}^{\dagger}\hat{c'}\rangle$ as given by $\propto
(n^{2}_{x}+n^{2}_{y} + 2 n_{x}n_{y}+ 2 n_{y}\sqrt{n_{x}n_{y}} cos
(2\theta_{1}+\phi_{m} + 2 n_{x}\sqrt{n_{x}n_{y}}
cos(2\theta_{1}+\phi_{m})$. It is more sensitive to the condition
$n_{x}=n_{y}$ or it will introduce more errors on Bob if the
condition $n_{x}=n_{y}$ is not fulfilled, even though its
contribution is in the order of mean photon number $n_{c}$ less than
the one photon number case ($|1\rangle_{x}|0\rangle_{y}$ and
$|0\rangle_{x}|1\rangle_{y}$). In other words, the existence of
multiple photons will introduce the percentage of errors in Bob even
without the presence of Eve. The percentage will also be the amount
information leaking to Eve in the photon-number splitting attack.

In intercept-resend attacks, Eve knows that the bit information is
not encoded in the state. He can perform the same measurement as Bob
did. Even though we assume that Eve break the security of the
$\alpha\eta$, Eve cannot resend a perfect copy of the original
signal photon because the input product of two mode weak coherent
states is a nonorthogonal state. For example, the input product
state $|\alpha\rangle_{x}|\beta\rangle_{y}$ of the beam splitter
(BS1) will have the output product state $(\frac{|\alpha\rangle_{x}+
ie^{\pm
i\phi_{m}}|\beta\rangle_{y}}{\sqrt{2}})_{2}(\frac{i|\alpha\rangle_{x}+e^{\pm
i\phi_{m}}|\beta\rangle_{y}}{\sqrt{2}})_{1}$, where the state
$(\frac{i|\alpha\rangle_{x}+ie^{\pm
i\phi_{m}}|\beta\rangle_{y}}{\sqrt{2}})_{2}\cdot\hat{e}_{45^{\circ}}$
is sent to Bob. The non-orthogonality of coherent states and the
$\phi_{m}=\pm 90^{\circ}$ cannot be measured in a single
measurement. Bob will reveal the present of Eve by sending part of
his observations (Step 3 and 4) to Alice. Let's say in the second
column in Fig.2, Alice knows that Bob is supposed to observe 'Yes'
with his guess through his projected angle $\theta_{2}=+45^{\circ}$.
Because of the imperfect copy from Eve, Bob might observe 'No' with
the $\theta_{2}=+45^{\circ}$, so Alice will aware of the error and
stop the transmission. There is no leakage of information to Eve
because Alice has not implemented Step 5 yet.

\section{CONCLUSION}

In conclusion, we have presented unique feature of a new scheme
based on intrinsic quantum correlation of weak coherent states for
generating key bits between Alice and Bob through high level
communication language. The scheme preserve the randomness of the
light source for doubling the communication distance, and also
increase the success rate of raw key generation compared to Ekert's
protocol, and double the sift key rate compared to weak coherent
light protocol.

\end{document}